\documentclass[preprint,12pt]{elsarticle}

\usepackage{amsmath}
\usepackage[utf8]{inputenc}
\usepackage{titlesec}
\biboptions{sort&compress}

\newcommand*{\defeq}{\mathrel{\rlap{%
                     \raisebox{0.3ex}{$\cdot$}}%
                     \raisebox{-0.3ex}{$\cdot$}}%
                     =}
\newcommand*{\reversedefeq}{=\mathrel{\rlap{%
                     \raisebox{0.3ex}{$\cdot$}}%
                     \raisebox{-0.3ex}{$\cdot$}}%
                     }
\usepackage{amssymb}

\newcounter{bla}

\journal{Computer Physics Communications}

\titleformat*{\subsection}{\normalsize\upshape\bf}

\newcommand{\cir}{\textasciicircum}

\begin{document}

\begin{frontmatter}



\title{RationalizeRoots: Software Package for the Rationalization of Square Roots}


\author[a,b]{Marco Besier}
\author[b]{Pascal Wasser}
\author[b]{Stefan Weinzierl} 

\address[a]{Institut f\"ur Mathematik, Johannes Gutenberg-Universit\"at Mainz, 55099 Mainz, Germany}
\address[b]{PRISMA Cluster of Excellence, Institut f\"ur Physik, Johannes Gutenberg-Universit\"at Mainz, 55099 Mainz, Germany}

\begin{abstract}
The computation of Feynman integrals often involves square roots. One way to obtain a solution in terms of multiple polylogarithms is to rationalize these square roots by a suitable variable change. We present a program that can be used to find such transformations. After an introduction to the theoretical background, we explain in detail how to use the program in practice.

\end{abstract}

\end{frontmatter}


\newpage
{\bf PROGRAM SUMMARY}
\par\vspace{\baselineskip}


\begin{small}
{\em Program title: RationalizeRoots}              
\par\vspace{\baselineskip}
{\em Licensing provisions: GNU General Public License 3}  
\par\vspace{\baselineskip}
{\em Programming language: Mathematica, Maple}    
 \par\vspace{\baselineskip}
{\em Program obtainable from: arXiv preprint server}
\par\vspace{\baselineskip}
{\em External programs required: Mathematica or Maple}
\par\vspace{\baselineskip}
{\em Nature of problem: Analytic solutions for Feynman integrals are critical for accurate theoretical predictions in high energy particle physics.
The computation of these integrals often involves square roots that need to be rationalized via suitable variable transformations.}
\par\vspace{\baselineskip}
{\em Solution method: Appropriate variable changes for given square roots are constructed by parametrizing algebraic hypersurfaces associated to these square roots by families of lines.}

\end{small}

\newpage
\section{Introduction}
\label{sec:Introduction}

The evaluation of Feynman integrals is one of the significant difficulties in the computation of multi-loop amplitudes in theoretical high energy particle physics.
A Feynman integral is a function of the space-time dimension $D$ and the kinematic variables.
Using dimensional regularization, one tries to write the integral as a Laurent series in $\epsilon=(4-D)/2$.
For a significant number of Feynman integrals, each term of their Laurent expansion can be written as a linear combination of multiple polylogarithms.
These functions are special iterated integrals with integration kernels of the form 

\begin{equation}
    \omega_j=\frac{dx}{x-z_j},
\end{equation}

\noindent where $z_j$ may depend on the kinematics, but is independent of the integration variable $x$.
In practice, however, we often encounter kernels like 

\begin{equation}
    \frac{dx}{\sqrt{(x-z_1)(x-z_2)}}.
\end{equation}

Therefore, we want to rationalize the square root that appears in the integration kernel to find a solution in terms of multiple polylogarithms.
Using partial fractioning, we can then express the integral in terms of the desired integration kernels plus trivial integrations.
\par\vspace{\baselineskip}
In recent years, the problem of rationalizing a given set of square roots by suitable variable changes played an important role in many physics applications \cite{Bourjaily:2018aeq,Becchetti:2017abb,Broadhurst:1993mw,Fleischer:1998nb,Davydychev:2000na,Jegerlehner:2002em,Kalmykov:2004kg,Aglietti:2004tq,Gehrmann:2018yef,Henn:2013woa,Lee:2017oca,Heller:2019gkq,Bork:2019aud,Abreu:2019rpt,Primo:2018zby}.
With this paper, we introduce a software package that implements the algorithm presented in \cite{Besier:2018jen} and solves the rationalization problem for a large class of square roots.
We provide two equivalent programs for the two computer algebra systems Mathematica and Maple.
We remark that Maple offers through the package ``algcurves'' the possibility to compute a rational parametrization of 
a genus zero curve based on the algorithm of \cite{vanHoeij:1997aab}.
This is limited to square roots in one variable. Our package goes beyond this limitation.
The paper is structured as follows:
section \ref{sec:TheoreticalBackground} covers the mathematical foundations and the rationalization methods that are used by the package.
Section \ref{sec:SetupAndDoc} shows how to load the package and documents the functions and options of the software.
Practical examples and applications are carried out in section \ref{sec:Applications}.
\newpage

\section{Theoretical Background}
\label{sec:TheoreticalBackground}

\subsection{Foundations}
\label{subsec:Foundations}
\textbf{Affine Hypersurfaces.}
An \textit{affine hypersurface} $V$ of degree $d$ is the zero set $\mathbb{V}(f)$ of a degree-$d$ polynomial $f\in \mathbb{C}[x_1,\ldots,x_n]$ in $n$ variables, embedded in $\mathbb{C}^n$: 
    
\begin{equation}
    V=\mathbb{V}(f)\subset \mathbb{C}^n.
\end{equation}
\par\vspace{\baselineskip}
\noindent Notice that the ambient space $\mathbb{C}^n$ is an essential part of our definition:
we explicitly spell out that the zero set of a polynomial in $n$ variables is always interpreted as a subset of the $n$\textit{-dimensional} space $\mathbb{C}^n$. 
Without specifying the embedding, the hypersurface $V=\mathbb{V}(x)$ might as well be viewed as the $y$-axis in $\mathbb{C}^2$.
From our definition, however, it is clear that we are talking about the one-point set $\{0\}\subset \mathbb{C}$ when writing $V=\mathbb{V}(x)$.
\par\vspace{\baselineskip}
\textbf{Projective Space.}
The \textit{projective $n$-space} $\mathbb{P}^n$ is the set of all complex lines through the origin in $\mathbb{C}^{n+1}$.
If $\sim$ denotes the equivalence relation of points lying on the same line through the origin, then

\begin{equation}
    \mathbb{P}^n=\frac{\mathbb{C}^{n+1}\backslash\{0\}}{\sim}.
\end{equation}
\par\vspace{\baselineskip}
\noindent Points in $\mathbb{P}^n$ are equivalence classes $[(x_0,\ldots,x_n)]=\{(\lambda x_0,\ldots,\lambda x_n)\}$, where $\lambda$ can be any non-zero complex number.
Notice that at least one of the coordinates $x_0,\ldots,x_n$ is non-zero.
We denote a point $p\in \mathbb{P}^n$ by one of its representatives.
To distinguish the class from its representative, we use square brackets rather than parenthesis and write colons between the coordinates of the representing point:

\begin{equation}
    [x_0:\ldots:x_n]\in \mathbb{P}^n.
\end{equation}

\noindent These \textit{homogeneous coordinates} emphasize that a point in $\mathbb{P}^n$ is only defined up to non-zero scalar multiple.
\par\vspace{\baselineskip}
\textbf{Points at Infinity.}
We may interpret $\mathbb{P}^n$ as the complex $n$-space $\mathbb{C}^n$ together with an ``infinitely distant point in every direction''.
To see this, consider the one-dimensional projective space $\mathbb{P}^1$, i.e., the set of all complex lines through the origin in $\mathbb{C}^2$.
By fixing a reference line in $\mathbb{C}^2$, i.e., a complex line not through the origin, we obtain a representative for each point $p\in \mathbb{P}^1$, namely the unique point where the reference line meets the line through the origin that defines $p$.
Only one point in $\mathbb{P}^1$ fails to have such a representative:
the point corresponding to the line through the origin that is parallel to the reference line.
We call this point the \textit{point at infinity}.
In summary, we have the identification
    
\begin{align}
\begin{split}
    \mathbb{P}^1&\rightarrow \mathbb{C}\cup \{\infty\},\\
    [x_0:x_1]&\mapsto
    \begin{cases}
    \frac{x_1}{x_0},\hspace{4pt}\text{for}\hspace{4pt}x_0\neq0,\\
    \infty,\hspace{4pt}\text{for}\hspace{4pt}x_0=0.
    \end{cases}
\end{split}
\end{align}

To take this idea one step further, consider the projective plane $\mathbb{P}^2$.
First, we fix a reference plane in $\mathbb{C}^3$ that does not pass through the origin.
Most points in $\mathbb{P}^2$ will have a unique representative on this reference plane.
The exceptions are the points corresponding to the lines through the origin that are parallel to the reference plane.
These \textit{points at infinity} form another copy of $\mathbb{P}^1$:

\begin{equation}
    \mathbb{P}^2=\mathbb{C}^2\cup \mathbb{P}^1.
\end{equation}

\noindent For instance, if $x_0,x_1,x_2$ are coordinates for $\mathbb{C}^{3}$ and we choose the reference plane $x_0=1$, we get the above equality by mapping the point $[x_0:x_1:x_2]$ to $(\frac{x_1}{x_0},\frac{x_2}{x_0})\in \mathbb{C}^2$ if $x_0\neq0$ and to $[x_1:x_2]\in \mathbb{P}^1$ if $x_0=0$.
\par\vspace{\baselineskip}
Generalizing this idea, we have 
\begin{align}
\begin{split}
    \mathbb{P}^n&=\mathbb{C}^n\cup \mathbb{P}^{n-1},\\
    [x_0:\ldots:x_n]&\mapsto
    \begin{cases}
    \left(\frac{x_1}{x_0},\ldots,\frac{x_n}{x_0}\right),\hspace{4pt}\text{for}\hspace{4pt}x_0\neq0,\\
    [x_1:\ldots:x_n],\hspace{4pt}\text{for}\hspace{4pt}x_0=0.
    \end{cases}
\end{split}
\end{align}

\noindent If $U_0$ is the set in $\mathbb{P}^n$ where the coordinate $x_0$ is non-zero, then we can identify $U_0$ with the hyperplane $x_0=1$ in $\mathbb{C}^{n+1}$ via

\begin{equation}
    [x_0:x_1:\ldots:x_n]=\left[1:\frac{x_1}{x_0}:\ldots:\frac{x_n}{x_0}\right]\mapsto \left(1,\frac{x_1}{x_0},\ldots,\frac{x_n}{x_0}\right).
\end{equation}

\noindent Thus, $U_0$ is a copy of $\mathbb{C}^n$ and we may think of it as the ``finite'' part of $\mathbb{P}^n$.
The remaining points, in which $x_0=0$, are the ``points at infinity''.
These are representatives of the lines through the origin in $\mathbb{C}^{n+1}$ that are parallel to the reference hyperplane $x_0=1$. They form an $(n-1)$-dimensional projective space $\mathbb{P}^{n-1}$.
\par\vspace{\baselineskip}
Notice that our choice of a reference hyperplane not containing the origin is arbitrary.
For instance, we could have chosen $x_i=1$ for any non-zero index $i\leq n$.
Therefore, what is ``finite'' and what is ``at infinity'' is a matter of perspective.
\par\vspace{\baselineskip}
\textbf{Projective Hypersurfaces.}
A polynomial $F\in \mathbb{C}[x_0,\ldots,x_n]$ is called \textit{homogeneous of degree $d$} if all its terms have the same degree $d$.
In particular, a degree-$d$ homogeneous polynomial satisfies

\begin{equation}
    F(\lambda x_0,\ldots,\lambda x_n)=\lambda^dF(x_0,\ldots,x_n),\hspace{4pt}\lambda\in \mathbb{C}.
\end{equation}

\noindent Notice that, if a point $(x_0,\ldots,x_n)\in \mathbb{C}^{n+1}$ is a zero of a homogeneous polynomial $F$, then every point $(\lambda x_0,\ldots,\lambda x_n)$ is a zero of $F$.
Thus, the zero set of $F$ is a union of complex lines through the origin in $\mathbb{C}^{n+1}$. 
\par\vspace{\baselineskip}
A \textit{projective hypersurface} is the set of zeros of a homogeneous polynomial $F\in \mathbb{C}[x_0,\ldots,x_n]$, embedded in $\mathbb{P}^n$:  
\begin{equation}
    V=\mathbb{V}(F)\subset \mathbb{P}^n.
\end{equation}

For example, the set $V=\mathbb{V}(x^2+y^2-z^2)\subset \mathbb{P}^2$ defines a projective hypersurface.
We can regard $V$ as the union of its intersections with the \textit{coordinate charts} of $\mathbb{P}^2$:
    
\begin{equation}
    V=(V\cap U_x)\cup (V\cap U_y)\cup (V\cap U_z).
\end{equation}
    
\noindent Notice that, on the set $U_z$ where $z\neq0$, the points of $V$ form a complex circle:
identifying $U_z$ with the hyperplane $z=1$, the curve in $U_z$ is defined by the vanishing of $x^2+y^2-1$.
Compare this to the sets where $x$ or $y$ are not zero: here, the defining equations are that of a hyperbola, namely $1+y^2-z^2=0$ and $x^2+1-z^2=0$, respectively.
As we see from this example, the intersection of any projective hypersurface with one of the affine coordinate charts of $\mathbb{P}^n$ defines an affine hypersurface, called an \textit{affine chart} of the projective hypersurface.
\par\vspace{\baselineskip}
\textbf{Projective Closure of an Affine Hypersurface.}
The \textit{projective closure} of an affine hypersurface $V=\mathbb{V}(f)\subset \mathbb{C}^n$ is the projective hypersurface $\overline{V}=\mathbb{V}(F)\subset \mathbb{P}^n$, where $F$ is the \textit{homogenization} of $f$.
We can \textit{homogenize} a degree-$d$ polynomial $f$ in $n$ variables to turn it into a degree-$d$ homogeneous polynomial $F$ in $n+1$ variables in the following way:
decompose $f$ into the sum of its \textit{homogeneous components} of various degrees, $f=g_0+\ldots+g_d$, where $g_i$ has degree $i$.
Notice that some $g_i$'s may be zero, but $g_d\neq0$.
The homogeneous component $g_d$ is already homogeneous of degree $d$.
The term $g_{d-1}\in \mathbb{C}[x_1,\ldots,x_n]$, however, is homogeneous of degree $d-1$.
To make it homogeneous of degree $d$ as well, we multiply by a new variable $x_0$ and obtain a polynomial $x_0g_{d-1}\in \mathbb{C}[x_0,\ldots,x_n]$.
In the same manner, every term $g_i$ can be made homogeneous of degree $d$ via multiplication by $x_0^{d-i}$.
The sum of these terms is the homogenization of $f$, a degree-$d$ homogeneous polynomial

\begin{equation}
    F=x_0^dg_{0}+x_0^{d-1}g_1+\ldots+g_d.
\end{equation}

\noindent We call $x_0$ the \textit{homogenizing variable}. 
Notice that the restriction of $F$ to the plane $x_0=1$ recovers the original polynomial $f$.
\par\vspace{\baselineskip}
As an example, consider the parabola $V=\mathbb{V}(y-x^2)\subset \mathbb{C}^2$.
The variables $x$ and $y$ are the affine coordinates for $V$, i.e., the coordinates of $\mathbb{C}^2$.
In the projective plane, however, we use homogeneous coordinates $x$, $y$, and $z$, thinking of $\mathbb{C}^2$ as the coordinate chart $U_z$ in which $z$ is non-zero.
We may identify $U_z$ with the plane $z=1$ in $\mathbb{C}^3$.
If we consider the parabola in $\mathbb{P}^2$, its points are the lines in $\mathbb{C}^3$ that connect the points on the parabola in the plane $z=1$ with the origin.
There is one line ``missing'' from the cone over the parabola:
the line $x=z=0$, i.e., the $y$-axis.
As a projective hypersurface, our parabola is defined by the equation $zy-x^2=0$.
It consists of the original parabola $y-x^2=0$ in the open set $U_z$, plus the ``infinitely distant point'' $[0:1:0]\in \mathbb{P}^2$.
\par\vspace{\baselineskip}
\textbf{Points of High Multiplicity.}
If $V=\mathbb{V}(f)$ is a hypersurface, affine or projective, a point $p\in V$ is said to be \textit{of multiplicity} $r\in \mathbb{N}$ if there exists at least one non-vanishing $r$-th partial derivative
    
\begin{equation}
    \frac{\partial ^{i_1+\ldots+i_n}f}{\partial x_1^{i_1}\cdots\partial x_n^{i_n}}(p)\neq0\hspace{4pt}\text{with}\hspace{4pt}i_1+\ldots+i_n=r
\end{equation}
    
\noindent and, at the same time, all lower-order partial derivatives vanish at $p$:
    
\begin{equation}
    \frac{\partial ^{i_1+\ldots+i_n}f}{\partial x_1^{i_1}\cdots\partial x_n^{i_n}}(p)=0\hspace{4pt}\text{with}\hspace{4pt}i_1+\ldots+i_n=k\hspace{4pt}\text{for all}\hspace{4pt}k=0,\ldots,r-1.
\end{equation}
    
\noindent We write $\text{mult}_p(V)=r$.
Points of multiplicity $1$ are called \textit{regular points} of $V$.
If $d$ denotes the degree of a given hypersurface, we will be particularly interested in the points of multiplicity $d-1$.
For the sake of simple language, we will speak of these points as $d-1$\textit{-points}, implicitly assuming that $d$ denotes the degree of the hypersurface under consideration.
\par\vspace{\baselineskip}
\textbf{Hypersurfaces Associated to Square Roots.}
Consider a square root $\sqrt{p/q}$ of a rational function, where $p$ and $q$ are polynomials.
We associate a hypersurface to this square root by naming it, e.g., denote it by $u$, squaring the resulting equation, and clearing the denominator.
Therefore, the associated hypersurface of $\sqrt{p/q}$ is given by $V=\mathbb{V}(q\cdot u^2-p)$.
\par\vspace{\baselineskip}
Notice that we can also associate a hypersurface to more general algebraic functions such as roots of degree greater than $2$ or nested roots.
For example, $V=\mathbb{V}(u^3-x^3-x^2)$ is associated to $\sqrt[\leftroot{0}\uproot{0}3]{x^3+x^2}$ and

\begin{equation}
    W=\mathbb{V}((u^2-x^2)^2-x^4-y^3)
\end{equation}

is associated to

\begin{equation}
    \sqrt{x^2+\sqrt{x^4+y^3}}.
\end{equation}

\subsection{Rationalization Algorithm}
\label{subsec:RationalizationAlgorithm}

\textbf{Rationalization of a Simple Root.}
We define a \textit{rational parametrization} of a hypersurface $V=\mathbb{V}(f)$ with $f\in \mathbb{C}[x_1,\ldots,x_n]$ as a set of rational functions $\phi_{x_1},\ldots,\phi_{x_n}$ that depend on $n-1$ new variables, say $t_1,\ldots,t_{n-1}$, and satisfy $f(\phi_{x_1},\ldots,\phi_{x_n})=0$.
\par\vspace{\baselineskip}
Given a square root, we can use a rational parametrization of its associated hypersurface to rationalize it.
Consider, for instance, $\sqrt{1-x^2}$.
The hypersurface associated to this square root is $V=\mathbb{V}(u^2+x^2-1)$, i.e., the unit circle in $\mathbb{C}^2$ with coordinates $u$ and $x$.
A rational parametrization of $V$ is, for example, given by

\begin{equation}
    (\phi_u(t),\phi_x(t))=\left(\frac{2t}{t^2+1},\frac{t^2-1}{t^2+1}\right).
\end{equation}

\noindent This tells us that changing variables like $x=(t^2-1)/(t^2+1)$ will transform $\sqrt{1-x^2}$ into the rational function $2t/(t^2+1)$. 
Indeed, we have

\begin{equation}
    \sqrt{1-x^2}=\sqrt{1-\left(\frac{t^2-1}{t^2+1}\right)^2}=\frac{2t}{t^2+1}.
\end{equation}

\textbf{The Algorithm.}
The rationalization method presented below is a slight generalization of the algorithm published in \cite{Besier:2018jen}.
The original formulation of the method did not include Step 5.
Instead, the variable $t_0$ was always assumed to be equal to $1$, and the possibility to make different choices when fixing $t_i$ was only given as a remark.
\par\vspace{\baselineskip}
\noindent \textbf{Input:} An irreducible degree-$d$ hypersurface $V$ whose projective closure $\overline{V}$ has at least one point of multiplicity $d-1$.
\par\vspace{\baselineskip}
\noindent \textbf{Output:} A rational parametrization of $V$.

\begin{itemize}
    \item[1.] Determine a point $p_0$ with $\text{mult}_{p_0}V=d-1$.
    \item[2.] If $p_0$ is not at infinity, go on with step 3. and 4. and finish with step 5.
              \par\vspace{\baselineskip}
              If $p_0$ is at infinity, consider another affine chart $V^\prime$ of the projective closure $\overline{V}$ in which $p_0$ is not at infinity, continue with steps 3., 4., 5., and finish with step 6.
    \item[3.] With $p_0=(a_0,\ldots,a_n)$, compute
              
              \begin{equation}
                  g(u,x_1,\ldots,x_n)\defeq f(u+a_0,x_1+a_1,\ldots,x_n+a_n)
              \end{equation}
              
              and write
    
              \begin{equation}
                 g(u,x_1,\ldots,x_n)=g_{d}(u,x_1,\ldots,x_n)+g_{d-1}(u,x_1,\ldots,x_n),
              \end{equation}
               
              where $g_{d}$ and $g_{d-1}$ are homogeneous components of degree $d$ and $d-1$.
    \item[4.] Return
    
              \begin{align}
              \begin{split}
                 \phi_u(t_0,\ldots,t_n)&=-t_0\frac{g_{d-1}(t_0,t_1,\ldots,t_n)}{g_d(t_0,t_1,\ldots,t_n)}+a_0,\\
                 \phi_{x_1}(t_0,\ldots,t_n)&=-t_1\frac{g_{d-1}(t_0,t_1,\ldots,t_n)}{g_d(t_0,t_1,\ldots,t_n)}+a_1,\\
                 &\vdots\\
                 \phi_{x_n}(t_1,\ldots,t_n)&=-t_n\frac{g_{d-1}(t_0,t_1,\ldots,t_n)}{g_d(t_0,t_1,\ldots,t_n)}+a_n.
              \end{split}
              \end{align} 
              
    \item[5.] For a single $i\in \{0,\ldots,n\}$, set $t_i=1$.
    \item[6.] Change coordinates to switch from $V^\prime$ to the original affine chart $V$.
\end{itemize}

\textbf{Example.}
Let us apply the algorithm to construct the aforementioned parametrization of the unit circle $V=\mathbb{V}(u^2+x^2-1)$.
\par\vspace{\baselineskip}
\noindent \textbf{Step 1.}
Because $\text{deg}(V)=2$, we can use any regular point of $V$ as our point of multiplicity $d-1$.
For instance, choose $p_0=(u_0,x_0)=(0,-1)$.
\par\vspace{\baselineskip}
\noindent \textbf{Step 2.}
$p_0$ is not a point at infinity, because it solves $u^2+x^2-1=0$.
\par\vspace{\baselineskip}
\noindent \textbf{Step 3.}
Define $g(u,x)\defeq f(u+0,x+(-1))=g_2(u,x)+g_1(u,x)$, where $g_2(u,x)=u^2+x^2$ and $g_1(u,x)=-2x$.
\par\vspace{\baselineskip}
\noindent \textbf{Step 4.}
Return
    
\begin{align}
    \begin{split}
    \phi_u(t_0,t_1)&=-t_0\frac{g_{1}(t_0,t_1)}{g_2(t_0,t_1)}+0,\\
    \phi_{x}(t_0,t_1)&=-t_1\frac{g_{1}(t_0,t_1)}{g_2(t_0,t_1)}+(-1).
    \end{split}
\end{align} 
    
\noindent \textbf{Step 5.}
Setting $t_0=1$, we obtain
    
\begin{align}
    \begin{split}
    \phi_u(t_1)&\defeq\phi_u(1,t_1)=-\frac{g_{1}(1,t_1)}{g_2(1,t_1)}=\frac{2t_1}{t_1^2+1},\\
    \phi_x(t_1)&\defeq\phi_{x}(1,t_1)=-t_1\frac{g_{1}(1,t_1)}{g_2(1,t_1)}-1=\frac{t_1^2-1}{t_1^2+1}.
    \end{split}
\end{align}

\subsection{F-Decomposition Theorem}
\label{subsec:FDecomposition}

\textbf{k-Homogenization.}
Let $k$ be a positive integer and $f\in \mathbb{C}[x_1,\ldots,x_n]$ a polynomial of degree $d$ with $d\leq k$.
The \textit{$k$-homogenization} of $f$ is a degree-$k$ homogeneous polynomial

\begin{equation}
    F(x_1,\ldots,x_n,z)\defeq z^k\cdot f\left(x_1/z,\ldots,x_n/z\right).
\end{equation}

\noindent For example, the $4$-homogenization of the polynomial $f(x_1,x_2)=x_1x_2$ is given by $F(x_1,x_2,z)=x_1x_2z^2$.
The $d$-homogenization of a degree-$d$ polynomial is the usual homogenization.
\par\vspace{\baselineskip}
In the following, we present an advanced rationalization method that is particularly useful in case the associated hypersurface of a given square root does not have a $d-1$-point.
The technique relies on the following theorem, proved in \cite{Besier:2018jen}.
\par\vspace{\baselineskip}
\noindent \textbf{Theorem.}
Let $V=\mathbb{V}(u^2-f_{\frac{d}{2}}^2+4f_{\frac{d}{2}+1}f_{\frac{d}{2}-1})$ be the hypersurface associated to

\begin{equation}
    \sqrt{f_{\frac{d}{2}}^2-4f_{\frac{d}{2}+1}f_{\frac{d}{2}-1}},
\end{equation}

\noindent where each $f_k\in \mathbb{C}[x_1,\ldots,x_n]$ is a polynomial of degree $\text{deg}(f_k)\leq k$.
Then, $V$ has a rational parametrization if $W=\mathbb{V}(F_{\frac{d}{2}+1}+F_{\frac{d}{2}}+F_{\frac{d}{2}-1})$ has a rational parametrization with $F_k$ being the $k$-homogenization of $f_k$ using the same homogenizing variable, say $z$, for each of the three homogenizations. 
\par\vspace{\baselineskip}
\noindent To be precise, if $(\phi_{x_1}^W,\ldots,\phi_{x_n}^W,\phi_{z}^W)$ is a rational parametrization of $W$, we obtain a rational parametrization of $V$ by defining

\begin{align}
\begin{split}
    \phi_u^V&\defeq 2\cdot\phi_z^W\cdot f_{\frac{d}{2}+1}\left(\phi_{x_1}^W/\phi_{z}^W,\ldots,\phi_{x_n}^W/\phi_{z}^W\right)+f_{\frac{d}{2}}\left(\phi_{x_1}^W/\phi_{z}^W,\ldots,\phi_{x_n}^W/\phi_{z}^W\right),\\
    \phi_{x_1}^V&\defeq \frac{\phi_{x_1}^W}{\phi_z^W},\\
    &\vdots\\
    \phi_{x_n}^V&\defeq \frac{\phi_{x_n}^W}{\phi_z^W}.
\end{split}
\end{align}

\noindent Notice that we use the letter $d$ in the index of the polynomials $f_k$ and $F_k$ because---in most physics applications--- this $d$ is equal to the degree of the square root argument considered in the theorem and therefore often equal to the degree of the hypersurface associated to the square root.
There are, however, some exceptions: 
for example, consider the following square root, which is relevant for the computation of hexagon integrals \cite{DelDuca:2011ne,Bourjaily:2018aeq}:

\begin{equation}
    \sqrt{(1-x_1-x_2-x_3)^2-4x_1x_2x_3}.
\end{equation}

\noindent Applying the theorem to this case, a possible choice of the polynomials $f_k$ is

\begin{align}
\begin{split}
        f_1(x_1,x_2,x_3)&=x_1,\\
        f_2(x_1,x_2,x_3)&=1-x_1-x_2-x_3,\\
        f_3(x_1,x_2,x_3)&=x_2x_3.
\end{split}
\end{align}

\noindent With this choice, we have $d=4$, while the degree of the square root argument is $3$.
In most other cases, however, $d$ will be equal to the degree of the polynomial under the square root.

\subsection{Sample Calculation}

This subsection presents a typical rationalization that requires all of the techniques discussed so far.
Consider the square root

\begin{equation}
    \sqrt{x^4+y^3}.
\end{equation}

The associated affine hypersurface is given by $V=\mathbb{V}(f)=\mathbb{V}(u^2-x^4-y^3)$.
Because $V$ has degree $4$, we need to find a point $p$ with $\text{mult}_pV=3$ to apply the rationalization algorithm.
Computing the partial derivatives of the homogenization $F$ of $f$, however, we see that $V$ does not have a point of multiplicity $3$---not even at infinity.
\par\vspace{\baselineskip}
We, therefore, use the $F$-decomposition: 
as a first step, we rewrite the square root as  

\begin{equation}
    \sqrt{x^4+y^3}=\sqrt{f_2^2-4f_3f_1}
\end{equation}

\noindent with 

\begin{equation}
    f_1(x,y)=-\frac{1}{4},\hspace{4pt}f_2(x,y)=x^2,\hspace{4pt}f_3(x,y)=y^3,
\end{equation}

\noindent and $k$-homogenizations

\begin{equation}
    F_1(x,y,z)=-\frac{1}{4}z,\hspace{4pt}F_2(x,y,z)=x^2,\hspace{4pt}F_3(x,y,z)=y^3.
\end{equation}

\noindent According to the theorem, $V$ has a rational parametrization if the hypersurface 

\begin{equation}
    W=\mathbb{V}(F_1+F_2+F_3)=\mathbb{V}(-z/4+x^2+y^3).
\end{equation}

\noindent has a rational parametrization.
Thus, we try to apply the algorithm to $W$:
\par\vspace{\baselineskip}
\noindent \textbf{Step 1.}
Because $\text{deg}(W)=3$, we need find a point of multiplicity $2$.
Looking at the partial derivatives of the $F_1+F_2+F_3$, we see that $W$ does not have such a point.
There is, however, a point of multiplicity $2$ at infinity.
We see this by considering the projective closure

\begin{equation}
    \overline{W}=\mathbb{V}(v^2F_1+vF_2+F_3).
\end{equation}

\noindent This projective hypersurface has a single point of multiplicity $2$, namely 

\begin{equation}
    p_0=[x_0:y_0:z_0:v_0]=[0:0:1:0].
\end{equation}
\noindent
\textbf{Step 2.}
Viewed from the affine chart $W$, $p_0$ is at infinity, because $v_0$ is zero.
Therefore, we have to consider a different affine chart $W^\prime$ of $\overline{W}$ in which $p_0$ is not at infinity.
In this particular example, we only have one choice, namely to consider the chart where $z=1$.
Switching from $\overline{W}$ to $W^\prime$ corresponds to a map
    
\begin{equation}
    [x:y:z:v]\mapsto \left(x/z,y/z,v/z\right)\reversedefeq\left(x^\prime,y^\prime,v^\prime\right).
\end{equation}
    
\noindent Under this mapping, $p_0\in \overline{W}$ is send to $p_0^\prime\defeq (0,0,0)\in W^\prime$.
The affine hypersurface $W^\prime$ is given by 

\begin{equation}
    W^\prime=\mathbb{V}\left(-\left(v^\prime\right)^2/4+v^\prime\left(x^\prime\right)^2+\left(y^\prime\right)^3\right).
\end{equation}
\par\vspace{\baselineskip}
\noindent
\textbf{Step 3.}
Define 

\begin{align}
\begin{split}
    g(x^\prime,y^\prime,v^\prime)&\defeq-\left(v^\prime+0\right)^2/4+\left(v^\prime+0\right)\left(x^\prime+0\right)^2+\left(y^\prime+0\right)^3\\
    &=g_3(x^\prime,y^\prime,v^\prime)+g_2(x^\prime,y^\prime,v^\prime),
\end{split}
\end{align}

\noindent where 

\begin{equation}
    g_3(x^\prime,y^\prime,v^\prime)\defeq v^\prime\left(x^\prime\right)^2+\left(y^\prime\right)^3\hspace{4pt}\text{and}\hspace{4pt}g_2(x^\prime,y^\prime,v^\prime)\defeq -\left(v^\prime\right)^2/4.
\end{equation}
\par\vspace{\baselineskip}
\noindent
\textbf{Step 4.}
Return
    
\begin{align}
\begin{split}
    \phi_{x^\prime}(t_0,t_1,t_2)&=-t_0\frac{g_{2}(t_0,t_1,t_2)}{g_3(t_0,t_1,t_2)}+0,\\
    \phi_{y^\prime}(t_0,t_1,t_2)&=-t_1\frac{g_{2}(t_0,t_1,t_2)}{g_3(t_0,t_1,t_2)}+0,\\
    \phi_{v^\prime}(t_0,t_1,t_2)&=-t_2\frac{g_{2}(t_0,t_1,t_2)}{g_3(t_0,t_1,t_2)}+0.
\end{split}
\end{align} 
    
\noindent\textbf{Step 5.}
Setting $t_0=1$, we obtain
    
\begin{align}
\begin{split}
    \phi_{x^\prime}(t_1,t_2)&\defeq\phi_{x^\prime}(1,t_1,t_2)=-\frac{g_{2}(1,t_1,t_2)}{g_3(1,t_1,t_2)}=\frac{t_2^2}{4(t_1^3+t_2)},\\
    \phi_{y^\prime}(t_1,t_2)&\defeq\phi_{y^\prime}(1,t_1,t_2)=-t_1\frac{g_{2}(1,t_1,t_2)}{g_3(1,t_1,t_2)}=\frac{t_1t_2^2}{4(t_1^3+t_2)},\\
    \phi_{v^\prime}(t_1,t_2)&\defeq\phi_{v^\prime}(1,t_1,t_2)=-t_2\frac{g_{3}(1,t_1,t_2)}{g_4(1,t_1,t_2)}=\frac{t_2^3}{4(t_1^3+t_2)}.
\end{split}
\end{align}
    
\noindent \textbf{Step 6.}
Finally, we have to translate the rational parametrization for $W^\prime$ into a rational parametrization for $W$.
To do this, we solve 
    
\begin{equation}
    \phi_{x^\prime}=\frac{\phi_{x}}{\phi_{z}},\hspace{4pt}\phi_{y^\prime}=\frac{\phi_{y}}{\phi_{z}},\hspace{4pt}\text{and}\hspace{4pt}\phi_{v^\prime}=\frac{\phi_{v}}{\phi_{z}}
\end{equation}
    
\noindent for $\phi_{x}$, $\phi_{y}$, and $\phi_{z}$ while putting $\phi_{v}=1$.
In this way, we obtain a rational parametrization of $W$ as
    
\begin{align}
\begin{split}
    \phi_x^W(t_1,t_2)&=\frac{1}{t_2},\\
    \phi_y^W(t_1,t_2)&=\frac{t_1}{t_2},\\
    \phi_z^W(t_1,t_2)&=\frac{4(t_1^3+t_2)}{t_2^3}.
\end{split}
\end{align}

\noindent As a last step, we use the $F$-decomposition theorem to obtain the change of variables that rationalizes $\sqrt{x^4+y^3}$:

\begin{align}
\begin{split}
    \phi_x^V(t_1,t_2)&=\frac{\phi_x^W(t_1,t_2)}{\phi_z^W(t_1,t_2)}=\frac{t_2^2}{4(t_1^3+t_2)},\\
    \phi_y^V(t_1,t_2)&=\frac{\phi_y^W(t_1,t_2)}{\phi_z^W(t_1,t_2)}=\frac{t_1t_2^2}{4(t_1^3+t_2)}.
\end{split}
\end{align}

\noindent Indeed, we have
    
\begin{equation}
    \sqrt{\left(\phi_x^V(t_1,t_2)\right)^4+\left(\phi_y^V(t_1,t_2)\right)^3}=\frac{t_2^3(2t_1^3+t_2)}{16(t_1^3+t_2)^2}.
\end{equation}

\newpage

\section{Setup and Documentation}
\label{sec:SetupAndDoc}

The {\tt RationalizeRoots} package comes in two versions:
one for Mathematica and one for Maple.
In the following, we give an overview of the functions and basic options.

\subsection{Mathematica}

The Mathematica program is loaded with the command
\par\vspace{\baselineskip}
{\tt Get["RationalizeRoots.m"]} \\
\par\vspace{\baselineskip}
The program provides the routines
\begin{itemize}
    \item {\tt ParametrizePolynomial[poly, options]}
    \begin{itemize}
        \item The input {\tt poly} is a (multivariate) polynomial.
        \item The output is a list of rational parametrizations for the hypersurface defined by {\tt poly}. 
        Each rational parametrization is given as a substitution list.
        By default, only one rational parametrization is returned.
        If no rational parametrization is found, the empty list is returned.
        \item Basic Options:
        \begin{itemize}
            \item {\tt Variables} $\rightarrow$ {\tt \{x1,x2,...\}}:
            Only the variables appearing in the list are considered as variables that can be reparametrized. All other variables are taken as parameters.
            In case this option is not specified, all variables appearing in {\tt poly} are considered as variables that can be reparametrized.
            \item {\tt OutputVariables} $\rightarrow$ {\tt \{y1,y2,...\}}: The variables appearing in the list are used as new variables.
            By default, {\tt t[1], t[2], ...} are used as new variables.
            \item {\tt MultipleSolutions} $\rightarrow$ {\tt True / False}: If true, a list of multiple rational parametrizations is returned.
            If false, the first rational parametrization found is returned. 
            The default value is false.
            \item {\tt GeneralC} $\rightarrow$ {\tt True / False}: If true, the rational parametrization may depend on free parameters {\tt C[1], C[2], ...}
            If false, a default value is substituted for all occurring free parameters.
            The default value is false.
            \item {\tt GeneralT} $\rightarrow$ {\tt True / False}: If true, the option skips step $5$ of the rationalization algorithm and leaves it to the user to set one of the new variables equal to one in the final change of variables.
            The default value is false.
            \item {\tt ForceFDecomposition} $\rightarrow$ {\tt True / False}: If true, try only the $F$-decomposi\-tion algorithm.
            The default value is false.
            \item {\tt FPolynomials} $\rightarrow$ {\tt\{f1,f2,f3\}}: Given the non-emp\-ty list\\
            {\tt \{f1,f2,f3\}}, assume that {\tt poly} is of the form {\tt u\cir2-f2\cir2+4f1f3} and use these polynomials for the $F$-decomposition.
            In case this option is not specified, a heuristic algorithm is used to find an $F$-decomposition.
        \end{itemize}
    \end{itemize}
    \item {\tt RationalizeRoot[root, options]}
        \begin{itemize}
        \item The input {\tt root} is of the form $R_1 \sqrt{R_2}$, where $R_1$ and $R_2$ are (multi-variate) rational functions.
        \item The output is a list of rationalizing variable changes.
        Each variable change is given as a substitution list.
        By default, only one variable change is returned.
        If no variable change is found, the empty list is returned.
        \item Basic Options:
        \begin{itemize}
            \item {\tt Variables}: As above.
            \item {\tt OutputVariables}: As above.
            \item {\tt MultipleSolutions}: As above.
            \item {\tt GeneralC}: As above.
            \item {\tt GeneralT}: As above.
            \item {\tt ForceFDecomposition}: As above.
            \item {\tt FPolynomials}: As above, but with the restriction that the input is assumed to be a square root of a polynomial $P$, which can be written as $\sqrt{P}=\sqrt{{\tt f2^2-4f1f3}}$.
        \end{itemize}
    \end{itemize}
\end{itemize}
\newpage

\subsection{Maple}

The Maple program is loaded with the command
\par\vspace{\baselineskip}
{\tt read "RationalizeRoots.mpl";} \\
\par\vspace{\baselineskip}
The program provides the routines
\begin{itemize}
    \item {\tt ParametrizePolynomial(poly, options)}
    \begin{itemize}
        \item The input {\tt poly} is a (multivariate) polynomial.
        \item The output is a list of rational parametrizations for the hypersurface defined by {\tt poly}. 
        Each rational parametrization is given as a substitution list.
        By default, only one rational parametrization is returned.
        If no rational parametrization is found, the empty list is returned.
        \item Basic Options:
        \begin{itemize}
            \item {\tt Variables = [x1,x2,...]}:
            Only the variables appearing in the list are considered as variables that can be reparametrized. 
            All other variables are taken as parameters.
            The default value is the empty list, in which case all variables appearing in {\tt poly} are considered as variables that can be reparametrized.
            \item {\tt OutputVariables = [y1,y2,...]}: The variables appearing in the list are used as new variables.
            By default, {\tt t\_1, t\_2, ...} are used as new variables.
            \item {\tt MultipleSolutions = true / false}: If true, a list of multiple rational parametrizations is returned.
            If false, the first rational parametrization found is returned.
            The default value is false.
            \item {\tt GeneralC = true / false}: If true, the rational parametrization may depend on free parameters {\tt C\_1, C\_2, ...}
            If false, a default value is substituted for all occurring free parameters.
            The default value is false.
            \item {\tt GeneralT = true / false}: If true, the option skips step $5$ of the rationalization algorithm and leaves it to the user to set one of the new variables equal to one in the final change of variables.
            The default value is false.
            \item {\tt ForceFDecomposition = true / false}: If true, try only the $F$-decomposi\-tion algorithm.
            The default value is false.
            \item {\tt FPolynomials = [f1,f2,f3]}: Given the non-emp\-ty list\\
            {\tt [f1,f2,f3]}, assume that {\tt poly} is of the form {\tt u\cir2-f2\cir2+4f1f3} and use these polynomials for the $F$-decomposition.
            The default value is the empty list, in which case a heuristic algorithm is used to find an $F$-decomposition.
        \end{itemize}
    \end{itemize}
    \item {\tt RationalizeRoot(root, options)}
        \begin{itemize}
        \item The input {\tt root} is of the form $R_1 \sqrt{R_2}$, where $R_1$ and $R_2$ are rational functions.
        \item The output is a list of rationalizing variable changes.
        Each variable change is given as a substitution list.
        By default, only one variable change is returned.
        If no variable change is found, the empty list is returned.
        \item Basic Options:
        \begin{itemize}
            \item {\tt Variables}: As above.
            \item {\tt OutputVariables}: As above.
            \item {\tt MultipleSolutions}: As above.
            \item {\tt GeneralC}: As above.
            \item {\tt GeneralT}: As above.
            \item {\tt ForceFDecomposition}: As above.
            \item {\tt FPolynomials}: As above, but with the restriction that the input is assumed to be a square root of a polynomial $P$, which can be written as $\sqrt{P}=\sqrt{{\tt f2^2-4f1f3}}$.
        \end{itemize}
    \end{itemize}
\end{itemize}

\newpage
\section{Applications}
\label{sec:Applications}

\subsection{RationalizeRoot}

When using the package for the first time, the {\tt RationalizeRoot} function is an excellent way to get started.
Without requiring any prior knowledge about the rationalization method, the user can provide a square root and obtain a variable change that turns this square root into a rational function.
For example, consider the square root $\sqrt{1-x^2-y^2}$. 
To find a rationalizing change of variables, we can apply the package as follows:
\par\vspace{\baselineskip}
In Mathematica:
\par\vspace{\baselineskip}
{\tt RationalizeRoot[Sqrt[1-x\cir2-y\cir2]]} \\
{\tt \{\{x$\rightarrow\frac{{\tt 2t[1]}}{{\tt 1+t[1]^2+t[2]^2}}$,y$\rightarrow-\frac{{\tt 1-t[1]^2+t[2]^2}}{{\tt 1+t[1]^2+t[2]^2}}$\}\}} 
\par\vspace{\baselineskip}
In Maple:
\par\vspace{\baselineskip}
{\tt RationalizeRoot(sqrt(1-x\cir2-y\cir2));} \\
{\tt [[x$=-\frac{{\tt 2\text{t\_1}}}{{\tt 1+\text{t\_1}^2+\text{t\_2}^2}}$,y$=-\frac{{\tt 2\text{t\_2}}}{{\tt 1+\text{t\_1}^2+\text{t\_2}^2}}$]]} 
\par\vspace{\baselineskip}
Both, the Mathematica and the Maple version of the program give valid transformations.
With these, we have $\sqrt{1-x^2-y^2}=2t_1t_2/(t_1^2+t_2^2+1)$ and $\sqrt{1-x^2-y^2}=(t_1^2+t_2^2-1)/(t_1^2+t_2^2+1)$, respectively.
\par\vspace{\baselineskip}
Although {\tt RationalizeRoot} is already quite powerful, it is considered a preliminary function.
For example, {\tt RationalizeRoot} will not rationalize nested square roots.
Using {\tt ParametrizePolynomial} instead, the user has more control over the hypersurface associated to the square root, which also allows for the rationalization of more general algebraic functions.
Advanced users should, therefore, work with the {\tt ParametrizePolynomial} function, which we will now present in detail.

\subsection{ParametrizePolynomial}
\label{subsec:ParametrizePolynomial}

As a first step, we demonstrate the basic usage of {\tt ParametrizePolynomial} using the square root $\sqrt{1-x^2-y^2}$.
Instead of the actual square root, we have to provide the defining polynomial of the associated hypersurface as input for the function:
\par\vspace{\baselineskip}
In Mathematica:
\par\vspace{\baselineskip}
{\tt ParametrizePolynomial[u\cir2+x\cir2+y\cir2-1]} \\
{\tt \{\{u$\rightarrow\frac{{\tt 2t[1]t[2]}}{{\tt 1+t[1]^2+t[2]^2}}$,x$\rightarrow\frac{{\tt 2t[1]}}{{\tt 1+t[1]^2+t[2]^2}}$,y$\rightarrow-\frac{{\tt 1-t[1]^2+t[2]^2}}{{\tt 1+t[1]^2+t[2]^2}}$\}\}} 
\par\vspace{\baselineskip}
In Maple:
\par\vspace{\baselineskip}
{\tt ParametrizePolynomial(u\cir2+x\cir2+y\cir2-1);} \\
{\tt [[u$=-\frac{{\tt 2}}{{\tt 1+\text{t\_1}^2+\text{t\_2}^2}}+1$,x$=-\frac{{\tt 2\text{t\_1}}}{{\tt 1+\text{t\_1}^2+\text{t\_2}^2}}$,y$=-\frac{{\tt 2\text{t\_2}}}{{\tt 1+\text{t\_1}^2+\text{t\_2}^2}}$]]} 
\par\vspace{\baselineskip}
We see that, in addition to the variable changes, the output also contains the expression of the rationalized square root up to sign.
Now that we understand the basic usage of {\tt ParametrizePoylnomial}, let us go through the different options of the function.
\par\vspace{\baselineskip}
\textbf{Variables.}
By default, {\tt ParametrizePolynomial} performs the transformation in all variables of the input.
Depending on the context, however, it can be advantageous to transform only a subset of the variables.
The {\tt Variables} option allows the user to specify which variables should be changed.
For example, consider the rationalization of $\sqrt{x+y+1}$.
On the one hand, we can rationalize using:
\par\vspace{\baselineskip}
In Mathematica:
\par\vspace{\baselineskip}
{\tt ParametrizePolynomial[u\cir2-x-y-1]} \\
{\tt \{\{u$\rightarrow\frac{{\tt 1+t[1]}}{{\tt t[2]}}$,x$\rightarrow\frac{{\tt 1+t[1]}}{{\tt t[2]}^2}$,y$\rightarrow-\frac{{\tt -t[1]}^2-{\tt t[1]+t[2]^2}}{{\tt t[2]}^2}$\}\}} 
\par\vspace{\baselineskip}
In Maple:
\par\vspace{\baselineskip}
{\tt ParametrizePolynomial(u\cir2-x-y-1);} \\
{\tt [[u$={\tt \text{t\_1} + \text{t\_2}}$,x$={\tt \text{t\_1} (\text{t\_1}+\text{t\_2})-1}$,y$={\tt \text{t\_2} (\text{t\_1}+\text{t\_2})}$]]} 
\par\vspace{\baselineskip}
On the other hand, we can use the {\tt Variables} option to only change variables in $y$:
\par\vspace{\baselineskip}
\newpage
In Mathematica:
\par\vspace{\baselineskip}
{\tt ParametrizePolynomial[u\cir2-x-y-1,Variables$\rightarrow$\{u,y\}]} \\
{\tt \{\{u$\rightarrow{\tt (1+x)t[1]}$,y$\rightarrow{\tt -1-x+t[1]^2+2xt[1]^2+x^2t[1]^2}$\}\}} 
\par\vspace{\baselineskip}  
In Maple:
\par\vspace{\baselineskip}
{\tt ParametrizePolynomial(u\cir2-x-y-1, Variables=[u,y]);} \\
{\tt [[u$={\tt \text{t\_1}}$,y$={\tt \text{t\_1}^2-x-1}$]]} 
\par\vspace{\baselineskip}
As we will see later, this option is particularly powerful when it comes to the simultaneous rationalization of multiple square roots.
We want to point out that, although the output obtained in this way is guaranteed to be rational in the new variables (in this case {\tt t[1]} or {\tt t\_1}), one might encounter new square roots that depend on the variables we viewed as parameters (in this case {\tt x}).
We will provide an in depth discussion of such an example in subsection \ref{subsec:RationalizingMultipleSquareRoots}.
\par\vspace{\baselineskip}
\textbf{OutputVariables.}
By default, the new variables of a transformation are called {\tt t[1]},{\tt t[2]}, \ldots, or {\tt t\_1},{\tt t\_2}, \ldots, as we have already seen above.
Using the option {\tt OutputVariables}, however, the user can specify the names of the new variables to be, for instance, $v$ and $w$:
\par\vspace{\baselineskip}
In Mathematica:
\par\vspace{\baselineskip}
{\tt ParametrizePolynomial[u\cir2+x\cir2+y\cir2-1, OutputVariables$\rightarrow$\{v,w\}]} \\
{\tt \{\{u$\rightarrow\frac{{\tt 2vw}}{{\tt 1+v^2+w^2}}$,x$\rightarrow\frac{{\tt 2v}}{{\tt 1+v^2+w^2}}$,y$\rightarrow-\frac{{\tt 1-v^2+w^2}}{{\tt 1+v^2+w^2}}$\}\}} 
\par\vspace{\baselineskip}
In Maple:
\par\vspace{\baselineskip}
{\tt ParametrizePolynomial(u\cir2+x\cir2+y\cir2-1, OutputVariables=[v,w]);} \\
{\tt [[u$=-\frac{{\tt 2}}{{\tt 1+v^2+w^2}}+1$,x$=-\frac{{\tt 2v}}{{\tt 1+v^2+w^2}}$,y$=-\frac{{\tt 2w}}{{\tt 1+v^2+w^2}}$]]} 
\par\vspace{\baselineskip}
This option is convenient when we apply the function iteratively to rationalize multiple square roots simultaneously.
\par\vspace{\baselineskip}
\textbf{MultipleSolutions.}
Setting {\tt MultipleSolutions} to {\tt True} provides the user with multiple rational parametrizations.
These parametrizations are obtained by applying the algorithm multiple times using all the different $d-1$-points across all affine charts of the projective closure of the given hypersurface. 
\par\vspace{\baselineskip}
\textbf{GeneralC.}
For some square roots, the associated hypersurface has an infinite number of $d-1$-points.
Consider, for instance, the square root $\sqrt{1-x^2}$, which is associated to the unit circle.
The unit circle is a hypersurface of degree 2.
Therefore, a $d-1$-point is given by any regular point.
The rational parametrization that the algorithm produces is, however, not independent of the choice of the $d-1$-point.
In fact, what point we choose will have an impact on the coefficients that we get in our final rationalizing change of variables.
To see this, we choose $(u_0,x_0)=(\sqrt{3}/2,1/2)$, instead of $(u_0,x_0)=(-1,0)$, as our $d-1$-point.
Making this choice produces rational parametrizations of the unit circle like

\begin{equation}
    \phi_u(t)=\frac{\sqrt{3}}{2}-\frac{\sqrt{3}+t}{t^2+1},\hspace{4pt}\phi_x(t)=\frac{1}{2}-\frac{t(\sqrt{3}+t)}{t^2+1}.
\end{equation}

The purpose of the {\tt GeneralC} option is to encode how the parametrization depends on the choice of the $d-1$-point---in case there are infinitely many of these points.
More precisely, if the option is enabled, the output will depend on free parameters {\tt C[1]}, {\tt C[2]}, etc., or {\tt C\_1}, {\tt C\_2}, etc.
Substituting concrete values for these parameters, the user is effectively fixing a $d-1$-point, in retrospect, which allows to produce a change of variables that is most suitable in the given context.
Applying the {\tt GeneralC} option to the unit circle, we get:
\par\vspace{\baselineskip}
In Mathematica:
\par\vspace{\baselineskip}
{\tt ParametrizePolynomial[u\cir2+x\cir2-1, GeneralC$\rightarrow$True]} \\
{\tt \{\{u$\rightarrow-\frac{{\tt C[2]-2C[1]t[1]+C[2]t[1]}^2}{{\tt C[1]-2C[2]t[1]+C[1]t[1]}^2}$,x$\rightarrow-\frac{\sqrt{{\tt C[1]}^2-{\tt C[2]}^2}-\sqrt{{\tt C[1]}^2-{\tt C[2]}^2}{\tt t[1]}^2}{{\tt C[1]-2C[2]t[1]+C[1]t[1]}^2}$\}\}} 
\par\vspace{\baselineskip}
In Maple:
\par\vspace{\baselineskip}
{\tt ParametrizePolynomial(u\cir2+x\cir2-1, GeneralC=true);} \\
{\tt [[u$=-\frac{{\tt 2\sqrt{1-\text{C\_2}^2}+2\text{C\_2}\text{t\_1} }}{{\tt 1+\text{t\_1}^2}}+\sqrt{1-\text{C\_2}^2}$,x$=-\frac{{\tt \text{t\_1}(2\sqrt{1-\text{C\_2}^2}+2\text{C\_2} \text{t\_1})}}{{\tt 1+\text{t\_1}^2}}+\text{C\_2}$]]} 
\par\vspace{\baselineskip}
In most cases, an integer choice of coordinates will produce the parametrizations that are least cluttered.
Therefore, whenever possible, the package chooses integer coordinates, if the {\tt GeneralC} option is not specified.
\par\vspace{\baselineskip}
Notice that the Mathematica version of this option differs slightly from the Maple version, due to the fact that different internal routines are used to find $d-1$-points.
As a consequence, the parametrizations obtained from the Mathematica program contain an additional free parameter with the only constraint that the user has to choose at least one of these parameters unequal to zero.
In Maple, on the other hand, the user is free to set all parameters equal to zero. 

\par\vspace{\baselineskip}
\textbf{GeneralT.}
The {\tt GeneralT} option skips step 5 of the rationalization algorithm and leaves it to the user to set one of the new variables $t_i$ equal to one in the final parametrization.
This has the advantage that one can spot what choice of $t_i=1$ produces the variable change that is most suitable in the user's context.
As an example, let us consider the hypersurface associated to $\sqrt{x^3+x^2}$.
Applying the {\tt GeneralT} option, we obtain:
\par\vspace{\baselineskip}
In Mathematica:
\par\vspace{\baselineskip}
{\tt ParametrizePolynomial[u\cir2-x\cir3-x\cir2, GeneralT$\rightarrow$True]} \\
{\tt \{\{u$\rightarrow\frac{{\tt t[1](-t[0]+t[1])(t[0]+t[1])}}{{\tt t[0]}^3}$,x$\rightarrow\frac{{\tt (-t[0]+t[1])(t[0]+t[1])}}{{\tt t[0]}^2}$\}\}} 
\par\vspace{\baselineskip}
In Maple:
\par\vspace{\baselineskip}
{\tt ParametrizePolynomial(u\cir2-x\cir3-x\cir2, GeneralT=true);} \\
{\tt [[u$=\frac{{\tt \text{t\_0}(\text{t\_0}^2-\text{t\_1}^2)}}{{\tt \text{t\_1}}^3}$,x$=\frac{{\tt \text{t\_0}^2-\text{t\_1}^2}}{{\tt \text{t\_1}}^2}$]]} 
\par\vspace{\baselineskip}
From this output, we see that we can simplify the variable change---in the sense that we avoid rational expressions---by choosing {\tt t[0]=1} (or {\tt t\_1=1}) instead of {\tt t[1]=1} (or {\tt t\_0=1}). 
Without setting {\tt GeneralT} to {\tt True}, the package would make a choice automatically, which does not necessarily lead to the best result possible.
\par\vspace{\baselineskip}
\textbf{ForceFDecomposition.}
Some square roots have the property that their associated hypersurface has a $d-1$-point and is, in addition, $F$-decomposable.
Consider, for instance, the following square root, which we already touched upon in subsection \ref{subsec:FDecomposition}:

\begin{equation}
    \sqrt{(1-x_1-x_2-x_3)^2-4x_1x_2x_3}.
\end{equation}

The associated hypersurface has several $d-1$-points, so the package will easily find multiple parametrizations.
In particular, it will not apply the $F$-decomposition theorem to generate these variable transformations.
We observe, however, that the square root is $F$-decomposable.
Therefore, we can use the {\tt ForceFDecomposition} option to force an application of the $F$-decomposition theorem.
This will give us variable changes that are, in general, different from the ones we get when not specifying the option.
In this way, we are able to produce even more variable changes for square roots of that type.
\par\vspace{\baselineskip}
\textbf{FPolynomials.}
Notice that, whenever we apply the $F$-decomposition theorem, we have a freedom in choosing $f_{\frac{d}{2}-1},f_{\frac{d}{2}}$, and $f_{\frac{d}{2}+1}$.
For the above square root, two appropriate choices would be:

\begin{align}
    \begin{split}
        &1.\hspace{4pt}f_1=1,\hspace{4pt}f_2=1-x_1-x_2-x_3,\hspace{4pt}f_3=x_1x_2x_3,\\
        &2. \hspace{4pt}f_1=x_1,\hspace{4pt}f_2=1-x_1-x_2-x_3,\hspace{4pt}f_3=x_2x_3.
    \end{split}
\end{align}

Making different choices for the $f_i$'s will result in different parametrizations.
Therefore, it can be useful to try different choices of the $f_i$'s to optimize the final variable transformation.
The user can specify a particular choice as follows:
if the input polynomial is of the form

\begin{equation}
    f\defeq u^2-f_{\frac{d}{2}}^2+4f_{\frac{d}{2}-1}f_{\frac{d}{2}+1}
\end{equation}

then the user has to provide the list 

\begin{equation}
    \{f_{\frac{d}{2}-1},f_{\frac{d}{2}},f_{\frac{d}{2}+1}\}.
\end{equation}

Notice that, in order to apply the $F$-decomposition with this particular choice of $f_i$'s, one has to set {\tt ForceFDecomposition} to {\tt True} in case $V\defeq\mathbb{V}(f)$ has a $d-1$-point.

\subsection{Simultaneous Rationalization of Multiple Square Roots}
\label{subsec:RationalizingMultipleSquareRoots}

From now on we will only present the {\tt Mathematica} commands and skip the corresponding {\tt Maple} commands for the sake of brevity.
\par\vspace{\baselineskip}
\textbf{A Simple Example.}
We begin the simultaneous rationalization of a set of square roots by virtue of a simple example.
Consider the two square roots:

\begin{equation}
    \{\sqrt{x+1},\sqrt{x+y+1}\}
\end{equation}

First, we rationalize $u\defeq\sqrt{x+y+1}$ via:
\par\vspace{\baselineskip}
{\tt ParametrizePolynomial[u\cir2-x-y-1,OutputVariables$\rightarrow$\{v,w\}]} \\
{\tt \{\{u$\rightarrow\frac{{\tt 1+v}}{{\tt w}}$,x$\rightarrow\frac{{\tt 1+v}}{{\tt w}^2}$,y$\rightarrow\frac{{\tt v(1+v)-w^2}}{{\tt w}^2}$\}\}}
\par\vspace{\baselineskip}  
This tells us that $\phi_x(v,w)\defeq (1+v)/w^2$ and $\phi_y(v,w)\defeq(v(1+v)-w^2)/w^2$ rationalize $\sqrt{x+y+1}$.
However, they do not rationalize $\sqrt{x+1}$.
In fact, under the above variable change, the square root $\sqrt{x+1}$ turns into

\begin{equation}
    r\defeq\sqrt{\frac{1+v}{w^2}+1}.
\end{equation}

To rationalize $\sqrt{x+1}$ and $\sqrt{x+y+1}$ simultaneously, we have to rationalize $r$ and compose the resulting transformation with the first variable change.
Therefore, the next step is to rationalize $r$ via:
\par\vspace{\baselineskip}
{\tt ParametrizePolynomial[r\cir2w\cir2-1-v-w\cir2]} \\
{\tt \{\{v$\rightarrow\frac{{\tt t[2]}^2{\tt -t[1]}^4{\tt -t[1]}^2{\tt t[2]}^2}{{\tt t[1]}^4}$,w$\rightarrow\frac{{\tt t[2]}}{{\tt t[1]}}$,r$\rightarrow\frac{{\tt 1}}{{\tt t[1]}}$\}\}}
\par\vspace{\baselineskip}  
Defining $\phi_v(t_1,t_2)\defeq(t_2^2-t_1^4-t_1^2t_2^2)/t_1^4$ and $\phi_w(t_1,t_2)\defeq t_2/t_1$, we conclude that $\varphi_x(t_1,t_2)\defeq\phi_x(\phi_v,\phi_w)$ and $\varphi_y(t_1,t_2)\defeq\phi_y(\phi_v,\phi_w)$ provide a change of variables that rationalizes $\sqrt{x+1}$ and $\sqrt{x+y+1}$ simultaneously.
Indeed, we find 

\begin{equation}
   \sqrt{\varphi_x+1}=\frac{1}{t_1}\hspace{4pt}\text{and}\hspace{4pt}\sqrt{\varphi_x+\varphi_y+1}=\frac{(t_1^2-1)t_2}{t_1^3}.
\end{equation}

Remarks:

\begin{itemize}
    \item In principle, the above method is not limited to a certain number of square roots.
    The problem of rationalizing multiple square roots simultaneously is, however, a very difficult one.
    From the authors' experience, the odds of finding a suitable transformation are relatively low when the number of square roots is significantly larger than the number of variables that are involved in the problem.
    \item Given a set of square roots, the change of variables that we find when using the above method is dependent on the ordering in which we rationalize the individual square roots.
    It might even be the case that choosing one ordering over the other is critical for the method to succeed at all.
    While not being a general rule, we found that starting with the rationalization of the most complicated square root and subsequently moving on to the simpler ones is often a good idea.
    \item Not all square roots can be rationalized, especially when one wants to rationalize many of them simultaneously.
    In fact, most square roots are not rationalizable, so it can be the case that the package does not find a rationalization, simply because there does not exist one.
    This is, for instance, the case when one encounters square roots associated to elliptic curves or K3 surfaces \cite{Laporta:2004rb,MullerStach:2011ru,Adams:2013kgc,Bloch:2013tra,Adams:2014vja,Adams:2015gva,Adams:2015ydq,Sogaard:2014jla,Bloch:2016izu,Remiddi:2016gno,Adams:2016xah,Bonciani:2016qxi,vonManteuffel:2017hms,Adams:2017ejb,Bogner:2017vim,Ablinger:2017bjx,Remiddi:2017har,Bourjaily:2017bsb,Hidding:2017jkk,Broedel:2017kkb,Broedel:2017siw,Broedel:2018iwv,Adams:2018yfj,Adams:2018bsn,Adams:2018kez,Henn:2013woa,Heller:2019gkq,Festi:092018,Bourjaily:2019hmc,Besier:2019hqd,Bogner:2019lfa,Bogner:2019vhn,Blumlein:2019svg,Bourjaily:2019jrk,Honemann:2018mrb,Bourjaily:2018yfy}.
    On the other hand, there exist some particular examples of square roots that are rationalizable, but not rationalizable using the package \cite{Besier:2018jen}. 
    These square roots are, however, very special and we are not aware of a single example in the context of high energy physics, where our package failed while another algorithm succeeded.
\end{itemize}

\textbf{Rationalization via Variables Option.}
The purpose of the following sample calculation is to show that the {\tt Variables} option can be critical when it comes to the rationalization of multiple square roots.
Suppose we want to rationalize 

\begin{equation}
    \{\sqrt{1-x^2},\sqrt{1-x^2-y^2}\}.
\end{equation}

Starting with the rationalization of the second square root, we find:
\par\vspace{\baselineskip}
{\tt ParametrizePolynomial[u\cir2+x\cir2+y\cir2-1,OutputVariables}$\rightarrow${\tt \{v,w\}]}  \\
{\tt \{\{u$\rightarrow\frac{{\tt 2vw}}{{\tt 1+v}^2{\tt +w}^2}$,x$\rightarrow\frac{{\tt 2v}}{{\tt 1+v}^2{\tt +w}^2}$,y$\rightarrow\frac{{\tt 2v}^2}{{\tt 1+v}^2{\tt +w}^2}-1$\}\}}
\par\vspace{\baselineskip}
The next step is to substitute the above expression for $x$ into $\sqrt{1-x^2}$ and try to rationalize the resulting square root.
We observe, however, that the package is not able to find a rationalization:
\par\vspace{\baselineskip}
\newpage
{\tt ParametrizePolynomial[r\cir2(1+v\cir2+w\cir2)\cir2+4v\cir2-(1+v\cir2+w\cir2)\cir2]}  \\
{\tt \{\}} 
\par\vspace{\baselineskip}
In such a case, the user might be tempted to think that a rationalization is not possible.
There is, however, a way in which we can still succeed, namely by using the {\tt Variables} option.
We start again by rationalizing $\sqrt{1-x^2-y^2}$, but this time we specify the {\tt Variables} option as follows:
\par\vspace{\baselineskip}
{\tt ParametrizePolynomial[u\cir2+x\cir2+y\cir2-1,Variables}$\rightarrow${\tt \{u,y\},}\\
\hspace{40pt} {\tt OutputVariables}$\rightarrow${\tt \{w\}]}  \\
{\tt \{\{u$\rightarrow\frac{{\tt 2w(x}^2{\tt -1)}}{{\tt (x}^2{\tt -1)w}^2{\tt -1}}$,y$\rightarrow\frac{\sqrt{{\tt 1-x}^2}{\tt ((x}^2{\tt -1)w}^2{\tt +1)}}{{\tt (x}^2{\tt -1)w}^2{\tt -1}}$\}\}}
\par\vspace{\baselineskip}
As already mentioned in subsection \ref{subsec:ParametrizePolynomial}, we see that the transformation is rational in the new variable $w$, but may contain square roots in the variable $x$ that we treated as a constant.
This square root in $x$ is, however, the second square root we wanted to rationalize originally.
Because the rationalization of $\sqrt{1-x^2-y^2}$ happened only via a change in $y$, the other original square root $\sqrt{1-x^2}$ does not change under this transformation.
Thus, we can simply rationalize the remaining square root via $x=(v^2-1)/(v^2+1)$ as in subsection \ref{subsec:RationalizationAlgorithm}. 
Substituting this expression for $x$ in the transformation of $y$ yields:

\begin{equation}
    x=\frac{v^2-1}{v^2+1},\hspace{4pt}y=-\frac{2v(1+v^4+v^2(2-4w^2))}{(1+v^2)(1+v^4+v^2(2+4w^2))}.
\end{equation}

Indeed, we can check that these substitutions turn the initial square roots into rational functions of $v$ and $w$:

\begin{equation}
    \sqrt{1-x^2}=\frac{2v}{v^2+1},\hspace{4pt}\sqrt{1-x^2-y^2}=\frac{8v^2w}{1+v^4+v^2(2+4w^2)}.
\end{equation}

\subsection{On the Role of Perfect Squares}

Especially in the process of rationalizing multiple square roots, it is often the case that one encounters square roots whose arguments contain factors that are perfect squares.
Notice that a rationalization of the square root without the perfect square factor already gives a rationalization of the square root that includes the perfect square factor.
For instance, consider the square root $\sqrt{x^3+x^2}=\sqrt{(x+1)x^2}$.
Since one of the factors of the argument is already a perfect square, it suffices to find a suitable variable change for the simpler square root $\sqrt{x+1}$, e.g., $x=t^2-1$, in order to rationalize $\sqrt{x^3+x^2}$.
\par\vspace{\baselineskip}
From the above example, one might be tempted to think that leaving out perfect squares is always a good idea.
This is, however, not always true.
\par\vspace{\baselineskip}
In fact, two cases can occur:

\begin{itemize}
    \item[1.] Leaving out a perfect square can make the rationalization procedure easier:
              \par\vspace{\baselineskip}
              The reader is invited to check that the package does not find a rational parametrization of $V=\mathbb{V}(u^2x^2-x^4-x^4y-xy^2-x^2 y^2)$, which is associated to the square root
              
              \begin{equation}
                  \sqrt{\frac{x^4 + x^4 y + x y^2 + x^2 y^2}{x^2}}.
              \end{equation}
              
              \par\vspace{\baselineskip}
              If we, however, leave out the perfect square in the denominator and instead consider 
              
              \begin{equation}
                  \sqrt{x^4 + x^4 y + x y^2 + x^2 y^2}
              \end{equation}
              
              with the associated hypersurface $W=\mathbb{V}(u^2-x^4-x^4y-xy^2-x^2 y^2)$, then the package will find a parametrization. This result can then, of course, also be used to rationalize the square root we wanted to rationalize in the first place.

    \item[2.] Leaving out a perfect square can make the rationalization procedure harder:
              \par\vspace{\baselineskip}
              Suppose we want to rationalize
              
              \begin{equation}
                  \sqrt{\frac{x^4+4x^2y^2+4}{4x^2}}.
              \end{equation}
              
              The reader can check that leaving out the perfect square in the denominator, i.e., considering 
              
              \begin{equation}
                  \sqrt{x^4+4x^2y^2+4}
              \end{equation}
              
              instead, leads to an associated hypersurface $V=\mathbb{V}(u^2-x^4-4x^2y^2-4)$, which does not have a single $d-1$-point.
              The package will still find a rational parametrization, but only after employing the $F$-decomposition theorem.
              \par\vspace{\baselineskip}
              If we, however, try to rationalize the original square root by considering $W=\mathbb{V}(4u^2x^2-x^4-4x^2y^2-4)$, we observe that $W$ has two $d-1$-points at infinity.
              Therefore, we can directly apply the algorithm so that, in this particular case, it is advantageous to keep the perfect square for the rationalization procedure.
\end{itemize}

From these two examples, we learn that it is a worthwhile exercise for the user to factor the perfect squares of the argument of the square root and try to find rationalizations while keeping and leaving out perfect squares as above.
We have seen that one can produce different, possibly refined, variable transformations with this strategy, which sometimes even allows for the rationalization of square roots whose associated hypersurface was---on first sight---not parametrizable by our methods.
\newpage

\section{Conclusions}
\label{sec:Conclusions}

In this paper, we introduced an implementation of the rationalization algorithm presented in \cite{Besier:2018jen}.
The problem of finding variable transformations that turn square roots into rational functions occurs in Feynman integral computations.
We covered the theoretical background and explained in detail how to load and use the software through various practical examples.
The program can rationalize many square roots occurring in high energy physics that admit a rationalization.
In the computation of recent perturbative corrections, e.g., mixed QCD-electroweak corrections for $H \rightarrow b\bar{b}$ \cite{Chaubey:2019lum}, the algorithm was used to construct rationalizing variable changes that were previously unknown.
We, therefore, expect this software package to be useful for future Feynman integral computations.
\par\vspace{\baselineskip}
\textbf{Acknowledgements}
\par\vspace{\baselineskip}
We thank Andreas von Manteuffel and Robert M. Schabinger for numerous discussions on the topic and for testing a preliminary version of the package early on. 
\newpage





\bibliographystyle{elsarticle-num}
\bibliography{bib}







\end{document}